\documentclass[12pt]{iopart}
\usepackage{iopams}
\expandafter\let\csname equation*\endcsname\relax
\expandafter\let\csname endequation*\endcsname\relax
\usepackage{amssymb,amsfonts,dsfont,mathrsfs,amsthm,mathtools}
\usepackage{cite}
\usepackage[english]{babel}
\newcommand{\diff}[1]{\text{d}#1}
\newcommand{\Diff}[1]{\text{D}#1}

\newcommand*{\diag}{\operatorname{diag}}

\newcommand{\Lie}{\mathcal{L}}
\newcommand{\Lag}{\mathscr{L}}

\begin{document}

\title{Symmetry algebra in gauge theories of gravity}
\author{Crist\'obal Corral \& Yuri Bonder}

\address{Instituto de Ciencias Nucleares, Universidad Nacional Aut\'onoma de M\'exico\\ Apartado Postal 70-543, Ciudad de M\'exico, 04510, M\'exico}
\ead{cristobal.corral@correo.nucleares.unam.mx; bonder@nucleares.unam.mx} 

\begin{abstract}
Diffeomorphisms and an internal symmetry (e.g., local Lorentz invariance) are typically regarded as the symmetries of any geometrical gravity theory, including general relativity. In the first-order formalism, diffeomorphisms can be thought of as a derived symmetry from the so-called local translations, which have improved properties. In this work, the algebra of an arbitrary internal symmetry and the local translations is obtained for a generic gauge theory of gravity, in any spacetime dimensions, and coupled to matter fields. It is shown that this algebra closes off shell suggesting that these symmetries form a larger gauge symmetry. In addition, a mechanism to find the symmetries of theories that have nondynamical fields is proposed. It turns out that the explicit form of the local translations depend on the internal symmetry and that the algebra of local translations and the internal group still closes off shell. As an example, the unimodular Einstein--Cartan theory in four spacetime dimensions, which is only invariant under volume preserving diffeomorphisms, is studied.
\end{abstract}

\maketitle

\section{Introduction}

Constructing a theory that reconciles quantum mechanics and general relativity is one of the greatest challenges in theoretical physics. Remarkably, an approach to quantization based on a gauge formulation of gravity has been successfully implemented in three spacetime dimensions~\cite{Witten:1988hc,Carlip:1998uc}. In this case, the theory is formulated in the grounds of the first-order formalism, namely, the triad and the connection are considered as independent fields, and it is (quasi)invariant under the local Poincar\'e group, when there is no cosmological constant, or (anti-) de Sitter [(A)dS] group when this constant is nonvanishing. The key observation is to identify the action principle as the Chern--Simons form of the (A)dS group~\cite{Witten:1988hc}, a method that can be generalized to arbitrary odd dimensions~\cite{Zanelli:2016cs}.

In arbitrary dimensions, the symmetries of the first-order formalism of gravity are usually taken to be local Lorentz transformations (LLT) and diffeomorphisms (Diff).  However, Diff is not a local symmetry in the sense that it relates tensors in different tangent spaces. Furthermore, it generates contracted Bianchi identities and off-shell conservation laws that are generically not covariant under LLT. Fortunately, this can be overcome by defining local translations (LT), which are local and fully covariant under LLT. In addition, it is strongly conjectured that the LT are a gauge symmetry in the full Hamiltonian sense~\cite{kiriushcheva2009hamiltonian}, and they are naturally associated with N\"other's theorem. What is more, the LLT and LT form an algebra that closes off shell, generalizing the Poincar\'e group in curved spacetimes~\cite{Hehl:1976kj,Blagojevic:2002du,Blagojevic:2013xpa}. 

Interestingly, the LT definition can be extended to cases where the gravitational theory is endowed with an arbitrary gauge symmetry~\cite{Obukhov:2006gea}. Moreover, it can be shown that infinitesimal Diff can be written as a linear combination of these generalized LT and the arbitrary gauge transformations (GT). This paper analyzes whether GT and LT form an algebra that closes off shell for a generic gauge theory of gravity. Additionally, theories for which the Diff or GT are explicitly broken are worked out. For the latter case an algorithm that selects the remaining symmetries is presented and the corresponding algebra is computed. It is shown that the algebra still closes off shell, strengthening the motivation that GT and LT form a gauge group. As an example, the unimodular theory of gravity with nontrivial torsion in four spacetime dimensions, namely, the unimodular Einstein--Cartan theory, is presented. 

The article is organized as follows: in Sec.~\ref{sec:PGT} the first-oder formalism is reviewed for an arbitrary internal group, in any dimension, and coupled to matter fields. The LT are defined in Sec.~\ref{sec:LT} where the algebra of this symmetry and the internal group is obtained. Sec.~\ref{sec:symmbreak} is devoted to analyze the situation where nondynamical fields break Diff and/or GT. Finally, the conclusions are presented in Sec.~\ref{sec:discussion}.

\section{Gauge theories of gravity}\label{sec:PGT}

The goal of this section is to review the first-order formalism of gravity for a generic internal group $\mathcal{G}$ with a Lie algebra $\mathfrak{g}$. The notation of Refs.~\cite{Hehl:1994ue,Obukhov:2006gea,Blagojevic:2013xpa} is followed, in particular, Latin and Greek indices are related to the internal group and spacetime, respectively. Given that the formalism involves spacetime $p$-forms, all spacetime indices are omitted, and that the dimensionality of the spacetime manifold $\mathcal{M}$ is $N$. The choice of $\mathcal{G}=GL(N,R)$ is of particular interest in metric-affine theories of gravity~\cite{Hehl:1994ue}. 

Consider a collection of $\mathfrak{g}$-valued differential forms $\Psi^A\in\Omega^p\left(\mathcal{M}\right)$ transforming infinitesimally as
\begin{align}
 \delta_{\rm GT}(\epsilon)\Psi^A = - \epsilon^I(x)\left(T_I\right)^{A}{}_B\Psi^B \equiv -\epsilon^{A}{}_B \Psi^B,
\end{align}
under the action of $\mathcal{G}$, where $I,J,K=1,\ldots,\dim\mathcal{G}$, while $\left(T_I\right)^{A}{}_B$ are the generators of the Lie algebra $\mathfrak{g}$, satisfying
\begin{align}
 \left[T_I,T_J\right]^{A}{}_B &= f_{IJK}\left(T^K\right)^{A}{}_B.
\end{align}
The collection of fields $\Psi^A$ includes all gravitational fields transforming in a nontrivial representation of $\mathcal{G}$ and it is also assumed that there exists a covariantly conserved rank-$2$ symmetric and gauge-invariant tensor $\pi_{AB}$, which can be used to lower internal indices, leading to $\epsilon_{AB} = - \epsilon_{BA}$. For example, when $\mathcal{G}=SO(1,N-1)$, one of the forms included in $\Psi^A$ is the $N$-dimensional vielbein $1$-form $e^a = e^{a}{}_\mu\diff{x^\mu}$, where, for convention, lower-case Latin characters denote Lorentz indices. As is customary, the vielbein is related to the spacetime metric $g_{\mu\nu}$ through $g_{\mu\nu} = \eta_{ab} e^{a}{}_\mu e^{b}{}_\nu$, where the $N$-dimensional Minkowski metric $\eta_{ab}=\diag\left(-,+,\cdots,+\right)$ is used. Still, $\Psi^A$ can include other gravitational degrees of freedom and arbitrary matter fields. Note that $\pi_{AB}$ is the generalization of $\eta_{ab}$ for an arbitrary $\mathcal{G}$.

On the other hand, the gauge connection $W^{A}{}_B$ is defined by the gauge-covariant derivative of $\Psi^A$, 
\begin{align}\label{derpsi}
 \Diff{\Psi^A} = \diff{\Psi^A} + W^{A}{}_B\wedge\Psi^B,
\end{align}
where $ \diff{}$ and $\wedge$ are, respectively, the exterior derivative and wedge product. Importantly, $ \Diff{\Psi^A}$ transforms covariantly under $\mathcal{G}$ provided that the gauge connection transforms as 
\begin{align}
\delta_{\rm GT}(\epsilon)W^{A}{}_B = \Diff{\epsilon^{A}{}_B}.
\end{align}
It it thus possible to summarize the transformation laws under GT by
\begin{align}
\label{GT}
 \mbox{GT} &= \begin{cases}
 \delta_{\rm GT}(\epsilon) W^{A}{}_B &= \Diff{\epsilon^{A}{}_B},\\
 \delta_{\rm GT}(\epsilon) \Psi^A &= -\epsilon^{A}{}_B \Psi^B.
 \end{cases}
\end{align}
In addition, the gauge curvature is defined as
\begin{align}\label{gaugecurvature}
 F^{AB} &= \diff{W^{AB}} + W^{A}{}_C\wedge W^{CB},
\end{align}
where the inverse of $\pi_{AB}$ has been used to raise internal group indices. It is easy to show that the gauge curvature satisfies the Bianchi identity $\Diff{F^{AB}} = 0$ and that the gauge connection and curvature are antisymmetric.

Now, (active) Diff generated by the vector field $\xi$ are implemented infinitesimally by the Lie derivative along $\xi$, i.e.,
\begin{align}
\label{Diff}
 \mbox{Diff} &= \begin{cases}
 \delta_{\rm Diff}(\xi) W^{A}{}_B &= \Lie_\xi W^{A}{}_B,\\
  \delta_{\rm Diff}(\xi) \Psi^A &= \Lie_\xi \Psi^A.
 \end{cases}\end{align}
According to Cartan's formula~\cite{Hehl:1994ue,Nakahara:2016}, when acting on a $p$-form $\alpha \in\Omega^p\left(\mathcal{M}\right)$, this derivative can be expressed by $\Lie_\xi\alpha = \diff{i_\xi}\alpha + i_\xi\diff{\alpha}$ where $i_{\xi}:\Omega^{p}\left(\mathcal{M}\right)\to\Omega^{p-1}\left(\mathcal{M}\right)$ is the inner contraction defined by
\begin{align}
 i_{\xi}\alpha = \frac{1}{\left(p-1\right)!}\xi^{\mu}\alpha_{\mu\nu_2\ldots\nu_p}\diff{x^{\nu_2}}\wedge\dots\wedge\diff{x^{\nu_p}}.
\end{align}

An action principle that is invariant under Diff and GT must be written as the spacetime integral of a gauge scalar $N$-form, $\Lag$, whose functional dependence on $W^{A}{}_B$ is through the covariant tensors $\Diff{\Psi^A}$ and $F^{AB}$, that is,
\begin{align}\label{fogaction}
 S\left[W^{AB},\Psi^A\right] &= \int\Lag\left[\Diff{\Psi^A},F^{AB},\Psi^A\right].
\end{align}
Notice that, in principle, $\Lag$ can include an exact form associated with the spacetime boundaries. An arbitrary variation of this action gives
\begin{align}\label{grav action var}
 \delta S &= \int\left( \delta W^{AB}\wedge\mathcal{E}_{AB}+\delta \Psi^A\wedge\mathcal{E}_A \right) - \int\diff{}\left(\delta W^{AB}\wedge H_{AB}+\delta \Psi^A\wedge H_A \right),
\end{align}
where 
\begin{align}
 \mathcal{E}_{AB} &= \Psi_{[A}\wedge H_{B]} - \Diff{H_{AB}},\label{EOM2}\\
 \mathcal{E}_A &= E_A  + (-1)^p\,\Diff{H_A},\label{EOM1}
\end{align}
and
\begin{align}\label{varderdef}
 H_A &= (-1)^{(N-p)(p+1)-p}\,\frac{\partial\Lag}{\partial\Diff{\Psi^A}}\bigg|_{\Psi,F}, \\
 H_{AB} &= - \frac{\partial\Lag}{\partial F^{AB}}\bigg|_{\Psi,\Diff{\Psi}},\\
  E_A &= (-1)^{p(N-p)}\frac{\partial\Lag}{\partial\Psi^A}\bigg|_{\Diff{\Psi},F}.
\end{align}
Note that Eq.~\eqref{EOM1} is actually a collection of equations since, recall, $\Psi^A$ represents a collection of forms. Observe that setting Eqs.~\eqref{EOM2} and~\eqref{EOM1} to zero gives the equations of motion for the collection of forms and the gauge connection, respectively. Thus, when one does not impose these equations to vanish, it is said that the results hold off shell.

If the action~\eqref{fogaction} is invariant under GT and Diff, the following N\"other identities hold:
\begin{align}
 \label{GTbianchigrav}
0 &= \Diff{\mathcal{E}_{AB}} - \Psi_{[A}\wedge \mathcal{E}_{B]},\\
0 &= (-1)^p\, i_A \Psi^B \wedge\Diff{\mathcal{E}_B} + i_A\Diff{\Psi^B}\wedge\mathcal{E}_B + i_A F^{BC}\wedge\mathcal{E}_{BC} \nonumber\\
& \quad - i_A W^{BC}\left( \Diff{\mathcal{E}_{BC}} - \Psi_{[B}\wedge \mathcal{E}_{C]}\right),\label{diffbianchigrav}
\end{align}
where\footnote{Note that $\xi^A = \theta^{A}{}_\mu \xi^\mu$, where $\theta^{A}{}_\mu$ are the components of the orthonormal frame.} $i_\xi = \xi^A i_A$ and the square brackets denote antisymmetrization with a $1/2$ factor. Equations \eqref{GTbianchigrav} and \eqref{diffbianchigrav} are obtained after inserting the corresponding transformation laws of $W^{AB}$ and $\Psi^A$ under GT [Eq.~\eqref{GT}] and Diff [Eq.~\eqref{Diff}] into the action variation~\eqref{grav action var}. In addition, the action~\eqref{fogaction} transforms as
\begin{align}
 \label{LLTgboundary}
 \delta_{\rm GT}(\epsilon) S &= 0,\\
\label{diffgboundary}
 \delta_{\rm Diff}(\xi) S &= \int\diff{}\left(i_\xi\Diff{\Psi^A}\wedge H_A - i_\xi\Psi^A\wedge E_A + i_\xi F^{AB}\wedge H_{AB}\right),
\end{align}
namely, it is invariant under GT and quasi-invariant under Diff, i.e., it transforms as a boundary term. These results are well known for generic $4$-dimensional gravity theories \cite{Kibble:1961ba,Sciama:1964wt,Hehl:1976kj} and in the context of $N$-dimensional Einstein--Cartan theory \cite{Kiriushcheva:2009ev}.

At this point it is possible to verify that, in general, the conservation law~\eqref{diffbianchigrav} depends on the connection and it is thus noncovariant. This is certainly an undesirable feature, which, together with the fact that Diff are not local, motivates the use of the LT. Of course, the term with the connection vanishes by virtue of Eq.~\eqref{GTbianchigrav}, which, however, could be modified at the quantum level (see, e.g., Ref.~\cite{Blagojevic2013,PhysRevD.96.044027}). Therefore, it is relevant to avoid introducing the connection-dependent term. The next section is devoted to present the LT and to build the algebra of such a symmetry and GT.

\section{Local translations and symmetry algebra}\label{sec:LT}

To modify the Diff so that the resulting transformations are covariant under GT one can simply take Cartan's formula and replace the exterior derivative with a gauge-covariant derivative. In fact, historically, LT were introduced precisely in this way (see Ref.~\cite{Hehl:1994ue} and references therein). Concretely, the fields $\Psi^A$ and $W^{AB}$ transform as
\begin{align}\label{LT}
 \text{LT} &= \begin{cases}
 \delta_{\rm LT}(\xi) W^{AB} = i_\xi F^{AB},\\
  \delta_{\rm LT}(\xi) \Psi^A\;\, = i_\xi\Diff{\Psi^A} + \Diff{i_\xi}\Psi^A.
 \end{cases}
\end{align}
Equivalently, one can show that the LT are a linear combination of Diff and a GT with the gauge parameter $\tilde{\epsilon}^{AB}=i_\xi W^{AB}$, that is 
\begin{align}\label{relation}
 \delta_{\rm Diff}(\xi) = \delta_{\rm LT}(\xi) + \delta_{\rm GT}(\tilde{\epsilon}).
\end{align}
Notice that $\tilde{\epsilon}^{AB}$ depends on the dynamical field $W^{AB}$. At first sight the fact that the gauge parameter depends on a dynamical variable seems unnatural. Nevertheless, it is analogous to what occurs in the well-known BRST symmetry. Furthermore, if the action principle~\eqref{fogaction} is invariant under GT and LT, then, the N\"other identities are
\begin{align}\label{GT1bianchigrav}
0 &= \Diff{\mathcal{E}_{AB}} -\Psi_{[A}\wedge \mathcal{E}_{B]},\\
\label{LTbianchigrav} 
 0 &= (-1)^p\, i_A \Psi^B \wedge\Diff{\mathcal{E}_B} + i_A\Diff{\Psi^B}\wedge\mathcal{E}_B + i_A F^{BC}\wedge\mathcal{E}_{BC},
\end{align}
respectively. Clearly, these expressions are gauge covariant, as expected since the LT are covariant by construction. Moreover, the N\"other current associated with GT is conserved on shell, while the corresponding current for the LT are generically not conserved. 

The algebra of GT and LT can be computed by acting on either $\Psi^A$ or $W^{AB}$. The method used to calculate the algebra is that of Ref.~\cite{Freedman:2012zz} where it is argued that only the dynamical fields transform, while the (arbitrary) gauge parameters remain unchanged. Importantly, this definition is compatible with the symmetry transformations of the fields obtained from the Poisson brackets with the conserved N\"other charges, which can be extended to the quantum realm.

The commutator of two GT acting on $\Psi^A$ reads
\begin{align}
 \left[\delta_{\rm GT}(\epsilon_1),\delta_{\rm GT}(\epsilon_2) \right] \Psi^A &= \delta_{\rm GT}(\epsilon_1)\left(-\epsilon_2^{A}{}_B\Psi^B \right) - \delta_{\rm GT}(\epsilon_2)\left(-\epsilon_1^{A}{}_B\Psi^B \right) \equiv -\epsilon_3^{A}{}_B\Psi^B,
\end{align}
where $\epsilon_{3}^{AB} = 2\epsilon_{1}^{[A}{}_C\epsilon_{2}^{|C|B]}$; this commutator acting on $W^{AB}$ takes the form
\begin{align}\notag
 \left[\delta_{\rm GT}(\epsilon_1),\delta_{\rm GT}(\epsilon_2)\right]W^{AB} &= \delta_{\rm GT}(\epsilon_1)\Diff{\epsilon_2^{AB}} - \delta_{\rm GT}(\epsilon_2)\Diff{\epsilon_1^{AB}} \\
  &= \left(\Diff{\epsilon_1^{A}{}_C}\right)\epsilon_2^{CB} + \left(\Diff{\epsilon_1^{B}{}_C}\right)\epsilon_2^{AC} - (1\leftrightarrow2)\equiv \Diff{\epsilon_3^{AB}}.
\end{align}
Hence,
\begin{align}
 \left[\delta_{\rm GT}(\epsilon_1),\delta_{\rm GT}(\epsilon_2)\right]  = \delta_{\rm GT}(\epsilon_3).
 \end{align}

The commutator of GT and LT acting on $\Psi^A$ gives
\begin{align}\notag
 \left[\delta_{\rm GT}(\epsilon),\delta_{\rm LT}(\xi) \right]\Psi^A &= \delta_{\rm GT}(\epsilon)\left[i_\xi\Diff{\Psi^A} + \Diff{i_\xi}\Psi^A \right] - \delta_{\rm LT}(\xi)\left[-\epsilon^{A}{}_B\Psi^B \right] \\
 \notag
 &= \xi^B\left[\epsilon^{C}{}_B i_C\Diff{\Psi^A} - \epsilon^{A}{}_C i_B\Diff{\Psi^C}\right] + \left(\Diff{\epsilon^{B}{}_C}\right)\xi^C i_B\Psi^A \\
 \notag
 &\quad +\Diff{\xi^B}\left[\epsilon^{C}{}_B i_C\Psi^A - \epsilon^{A}{}_C i_B\Psi^C \right] + \xi^B\left[\epsilon^{C}{}_B\Diff{i_C\Psi^A} - \epsilon^{A}{}_C\Diff{i_B\Psi^C} \right] \\
 &\quad + \epsilon^{A}{}_B\left[i_\xi\Diff{\Psi^B} + \Diff{i_\xi}\Psi^B \right]
 \equiv i_{\tilde{\xi}}\Diff{\Psi^A} + \Diff{i_{\tilde{\xi}}}\Psi^A,
\end{align}
where $\tilde{\xi}^A = \epsilon^{A}{}_B\xi^B$. Analogously, this commutator, when acting on $W^{AB}$, yields
\begin{align}\notag
 \left[\delta_{\rm GT}(\epsilon),\delta_{\rm LT}(\xi) \right]W^{AB} &= \delta_{\rm GT}(\epsilon)i_\xi F^{AB} - \delta_{\rm LT}(\xi)\Diff{\epsilon^{AB}}\\
 \notag
 &= \xi^C\left[\epsilon^{D}{}_C i_D F^{AB} - \epsilon^{A}{}_D i_C F^{DB} - \epsilon^{B}{}_C i_C F^{AD} \right]\\ 
 &\quad - i_\xi F^{A}{}_C \epsilon^{CB} - i_\xi F^{B}{}_C \epsilon^{AC} \equiv i_{\tilde{\xi}} F^{AB}.
\end{align}
These results can be summarized as
\begin{align}
 \left[\delta_{\rm GT}(\epsilon),\delta_{\rm LT}(\xi)\right] &= \delta_{\rm LT}(\tilde{\xi}). \end{align}

Finally, the commutator of two LT acting on $\Psi^A$ gives
\begin{align}
 \left[\delta_{\rm LT}(\xi_1),\delta_{\rm LT}(\xi_2) \right]\Psi^A &= \delta_{\rm LT}(\xi_1)\left[i_{\xi_2}\Diff{\Psi^A} + \Diff{i_{\xi_2}}\Psi^A \right] - (1\leftrightarrow2) \nonumber\\
 &= \xi_2^B\left[\Lie_{\xi_1}\left(i_B\Diff{\Psi^A}\right) - \left(i_{\xi_1}W^{C}{}_B\right) i_C\Diff{\Psi^A} + \left(i_{\xi_1}W^{A}{}_C\right) i_B\Diff{\Psi^C}  \right]\nonumber\\ 
 &\quad + \left(i_{\xi_1}F^{B}{}_C\right)\xi_2^Ci_B\Psi^A + \Diff{\xi_2^B}\big[\Lie_{\xi_1}\left(i_B\Psi^A\right) - \left(i_{\xi_1}W^{C}{}_B\right) i_C\Psi^A\nonumber\\
 &\quad + \left(i_{\xi_1}W^{A}{}_C \right)i_B\Psi^C \big] +  \xi_2^B\big[\Lie_{\xi_1}\left(\Diff{i_B\Psi^A} \right) - \left(i_{\xi_1}W^{C}{}_B\right)\Diff{i_C\Psi^A} \nonumber\\ 
 &\quad + \left(i_{\xi_1}W^{A}{}_C\right)\Diff{i_B\Psi^C} \big] - (1\leftrightarrow2).
\end{align}
Using the Leibniz rule for the Lie derivative and the identity
\begin{align}\notag
 -\left(i_{\xi_1}i_{\xi_2}F^{A}{}_B\right)\Psi^B + \Diff{}\left(\left[i_{\xi_1},\Lie_{\xi_2}\right]\Psi^A \right)  &= i_{\xi_1}\Diff{i_{\xi_2}}\Diff{\Psi^A}  + i_{\xi_1}\Diff{}\Diff{i_{\xi_2}}\Psi^A + \Diff{i_{\xi_1}i_{\xi_2}}\Diff{\Psi^A} \\
 &\quad + \Diff{i_{\xi_1}}\Diff{i_{\xi_2}}\Psi^A - (1\leftrightarrow2),
\end{align}
one obtains
\begin{align}
 \left[\delta_{\rm LT}(\xi_1),\delta_{\rm}(\xi_2) \right] \Psi^A &\equiv  i_{\bar{\xi}}\Diff{\Psi^A} + \Diff{i_{\bar{\xi}}}\Psi^A - \bar{\epsilon}^{A}{}_B \Psi^B,
\end{align}
where $\bar{\xi}^A = \Lie_{\xi_1}\xi_2^A + i_{\xi_2}\Diff{\xi_1^A} - i_{\xi_1}\Diff{\xi_2^A}$ and $\bar{\epsilon}^{AB} = i_{\xi_1}i_{\xi_2}F^{AB}$. Observe that $\bar{\xi}^A = i_{\xi_1}i_{\xi_2}T^A$ where $T^A$ is the torsion of the gauge connection $W^{AB}$. In a similar way, the commutator of two LT acting on $W^{AB}$ is
\begin{align}
 \left[\delta_{\rm LT}(\xi_1),\delta_{\rm LT}(\xi_2) \right]W^{AB} &= \delta_{\rm LT}(\xi_1)i_{\xi_2}F^{AB} - \delta_{\rm LT}(\xi_2)i_{\xi_1}F^{AB}\nonumber \\
 &= \xi_2^C\big[\Lie_{\xi_1}\left(i_C F^{AB} \right) - \left(i_{\xi_1}W^{D}{}_C\right) i_D F^{AB} \nonumber\\ &\quad + \left(i_{\xi_1}W^{A}{}_D \right)i_C F^{DB} + \left(i_{\xi_1}W^{B}{}_D\right)i_C F^{AD}\big] - (1\leftrightarrow2).
\end{align}
In this case one needs to use
\begin{align}
 \Lie_{\xi_2}i_{\xi_1}F^{AB} &= i_{\xi_2}\Lie_{\xi_1} F^{AB} + \left(i_{\xi_1}W^{A}{}_C \right)i_{\xi_2}F^{CB} - \left(i_{\xi_2}W^{A}{}_C \right)i_{\xi_1}F^{CB} \nonumber\\
 &\quad + \left(i_{\xi_1} W^{B}{}_C \right) i_{\xi_2} F^{AC} - \left(i_{\xi_2}W^{B}{}_C\right)i_{\xi_1}F^{AC} + \Diff{}\left(i_{\xi_2}i_{\xi_1}F^{AB} \right),
\end{align}
and the relation $\left[\Lie_{\xi_1},i_{\xi_2}\right]\alpha = i_{\left[\xi_1,\xi_2 \right]}\alpha$, where $\alpha$ is a generic form, to obtain
\begin{align}
 \left[\delta_{\rm LT}(\xi_1),\delta_{\rm LT}(\xi_2) \right]W^{AB} &= \left(\Lie_{\xi_1}\xi_2^C - i_{\xi_1}\Diff{\xi_2^C} + i_{\xi_2}\Diff{\xi_1^C} \right)i_C F^{AB} + \Diff{}\left(i_{\xi_1}i_{\xi_2}F^{AB} \right) \nonumber\\
 &= i_{\bar{\xi}}F^{AB} + \Diff{\bar{\epsilon}^{AB}}.
\end{align}
These expressions can be condensed into
\begin{align}
 \left[\delta_{\rm LT}(\xi_1),\delta_{\rm LT}(\xi_2)\right] &= \delta_{\rm GT}(\bar{\epsilon}) + \delta_{\rm LT}(\bar{\xi}).
\end{align}

In conclusion, the algebra of GT and LT is
\begin{align}
 \left[\delta_{\rm GT}(\epsilon_1),\delta_{\rm GT}(\epsilon_2)\right] &= \delta_{\rm GT}(\epsilon_3),\\
 \left[\delta_{\rm GT}(\epsilon),\delta_{\rm LT}(\xi)\right] &= \delta_{\rm LT}(\tilde{\xi}), \\
 \label{LTLT}
 \left[\delta_{\rm LT}(\xi_1),\delta_{\rm LT}(\xi_2)\right] &= \delta_{\rm GT}(\bar{\epsilon}) + \delta_{\rm LT}(\bar{\xi}), 
\end{align}
where $\epsilon_{3}^{AB} =2\epsilon_{1}^{[A}{}_C\epsilon_{2}^{|C|B]}$, $\tilde{\xi}^A = \epsilon^{A}{}_B\xi^B$, $\bar{\epsilon}^{AB} = i_{\xi_1}i_{\xi_2}F^{AB}$, $\bar{\xi}^A = \left[\xi_1,\xi_2\right]^A + i_{\xi_2}\Diff{\xi_1^A} - i_{\xi_1}\Diff{\xi_2^A}$. Interestingly, this algebra closes off shell. Also, it should be emphasized that, in contrast to what occurs in three-dimensional vacuum general relativity, this algebra is not local Poincar\'e since the LT do not commute among themselves. The analysis presented here has been carried out in very general grounds: it applies to gravity gauge theories in arbitrary spacetime dimensions, with an arbitrary internal symmetry group, and possibly with additional gravitational degrees of freedom and matter fields. The next section is devoted to study the case of gravitational theories with broken symmetries.

\section{Theories with explicit symmetry breaking\label{sec:symmbreak}}

The relation between LT, Diff, and GT in Eq.~\eqref{relation} implies that any theory that is invariant under two of these symmetries, must be invariant under the third. Moreover, in these cases it is easy to find the action of all the symmetries on the dynamical fields. Conversely, if a theory (partially) breaks these symmetries by the presence of nondynamical fields (which do not transform), then, the other symmetries must be affected. Of course, to study the algebra of GT and LT in this context, one first needs to find the corresponding field transformations. For this purpose an algorithm that selects the symmetries, and which can be applied to theories with explicit symmetry breaking, is presented.

This algorithm is inspired by Ref.~\cite{Montesinos:2017epa,MercedMatter,MercedLovelock} and, for a generic theory described by an action
\begin{align}\label{totalaction}
 S[W^{AB},\Psi^A;\Phi] &= \int \Lag\left[\Diff{\Psi^A},F^{AB},\Psi^A;\Phi\right],
 \end{align}
 where $\Phi$ is a collection of nondynamical forms, it has five basics steps:
 \begin{enumerate}
 \item Perform an arbitrary variation of the action~\eqref{totalaction}
\begin{align}\label{vartotalaction}
 \delta S &= \int\left( \delta W^{AB}\wedge \mathcal{E}_{AB} +\delta \Psi^A\wedge \mathcal{E}_A  \right) ,
 \end{align}
where the boundary terms are neglected from now on. Note that $\mathcal{E}_{AB}$ and $\mathcal{E}_{A}$ can depend on $\Phi$.
\item Apply the gauge-covariant derivative on $\mathcal{E}_{AB}$ and $\mathcal{E}_{A}$ and write the resulting expression as linear combinations of $\mathcal{E}_{AB}$ and $\mathcal{E}_{A}$ plus ``symmetry breaking terms''. That is, find the forms $ {\alpha_{AB}}^{CD}$, ${ \beta_{AB}}^C$, $\gamma_{AB}$,  $ {\chi_{A}}^{CD}$, ${\zeta_A}^B$, and $\rho_A $ such that
\begin{align}
\label{DFab}
\Diff{\mathcal{E}_{AB}} &= {\alpha_{AB}}^{CD}\wedge \mathcal{E}_{CD} +{ \beta_{AB}}^C\wedge  \mathcal{E}_C + \gamma_{AB},\\
\label{DFa}
\Diff{\mathcal{E}_A} &=  {\chi_{A}}^{BC}\wedge \mathcal{E}_{BC}+{\zeta_A}^B\wedge \mathcal{E}_B + \rho_A ,
\end{align}
where $\gamma_{AB}$ and $\rho_{A}$ are the symmetry breaking terms (this terminology is clarified below).
\item Contract Eqs.~\eqref{DFab} and~\eqref{DFa} with the gauge parameters $\epsilon^{AB}$ and $\xi^A$, respectively, and use the Leibniz rule to take these equations into the form
\begin{align}
 \label{DFabepsab}
\diff{}\left(\epsilon^{AB} \mathcal{E}_{AB}\right) &= \left(\Diff{\epsilon^{AB}} +\epsilon^{CD} {\alpha_{CD}}^{AB}\right)\wedge\ \mathcal{E}_{AB} +\epsilon^{BC}{\beta_{BC}}^A \wedge \mathcal{E}_A +  \epsilon^{AB} \gamma_{AB},\\
\label{DFaxia}
 \diff{}\left(\xi^A \mathcal{E}_A\right) &=  {\xi^C\chi_{C}}^{AB}\wedge\ \mathcal{E}_{AB} +\left(\Diff{\xi^A} +\xi^B {\zeta_B}^A \right) \wedge\ \mathcal{E}_A + \xi^A\,\rho_A.
 \end{align}
\item Integrate Eqs.~\eqref{DFabepsab} and~\eqref{DFaxia} over $\mathcal{M}$ to obtain
\begin{align}\label{gaugelawlt} 
0&= \int\left[ \left(\Diff{\epsilon^{AB}} +\epsilon^{CD} {\alpha_{CD}}^{AB}\right)\wedge\ \mathcal{E}_{AB} +\epsilon^{BC}{\beta_{BC}}^A \wedge \mathcal{E}_A +  \epsilon^{AB} \gamma_{AB}\right],\\
\label{tranlawlt}
0 &= \int\left[ {\xi^C\chi_{C}}^{AB}\wedge\ \mathcal{E}_{AB} +\left(\Diff{\xi^A} +\xi^B {\zeta_B}^A \right) \wedge\ \mathcal{E}_A +\xi^A\,\rho_A\right],
 \end{align}
 where boundary terms have been neglected.
\item If one can find a family of gauge parameters $\epsilon^{AB}$ ($\xi^A$) such that $\epsilon^{AB}\gamma_{AB}=0$ ($\xi^A \rho_A=0$), then, by comparing with the action variation \eqref{vartotalaction}, it is clear that there is a remaining GT (LT) symmetry, which is pointed out in the following expressions:
\begin{align}
0&= \int\Big[ \underbrace{\left(\Diff{\epsilon^{AB}} +\epsilon^{CD} {\alpha_{CD}}^{AB}\right)}_{\delta_{\rm GT}(\epsilon) W^{AB}}\wedge\, \mathcal{E}_{AB} +\underbrace{\epsilon^{BC}{\beta_{BC}}^A }_{\delta_{\rm GT}(\epsilon)\Psi^A} \wedge\, \mathcal{E}_A \Big],\\
0 &= \int\Big[\underbrace{ {\xi^C\chi_{C}}^{AB}}_{\delta_{\rm LT}(\xi) W^{AB}}\wedge\, \mathcal{E}_{AB} +\underbrace{\left(\Diff{\xi^A} +\xi^B {\zeta_B}^A \right)}_{\delta_{\rm LT}(\xi)\Psi^A} \wedge\, \mathcal{E}_A \Big],
 \end{align}
where, recall, the gauge parameters are subject to the constraints that they vanish when contracted with the symmetry breaking terms. Clearly, if there are no restricted gauge parameters for which the  symmetry breaking terms vanish, then, the theory is simply not invariant under the corresponding symmetry. (For an example of a theory that has no remaining symmetries see Ref.~\cite{BonderSymmetry}). 
\end{enumerate}
Note that, in theories that are invariant under GT or/and LT, the symmetry breaking terms do not appear, thus, there are no constraints on the gauge parameters. This justifies the terminology for the symmetry breaking terms. Moreover, once the symmetry transformations are known, it is possible to verify if they form a closed algebra; this is done next in a concrete example.

\subsection{Unimodular Einstein--Cartan}

In this subsection, the explicit form of the LT and the algebra with the internal symmetries is analyzed in the unimodular Einstein--Cartan theory of gravity. This example shows how the algorithm described above can be applied in a theory that is invariant under a smaller set of symmetries and how to compute the algebra in these cases.

Unimodular gravity is an appealing proposal to address the cosmological constant problem~\cite{Einstein:1919gv,Anderson:1971pn,vanderBij:1981ym,Buchmuller:1988wx,Unruh:1988in,Henneaux:1989zc,Ng:1990xz,finkelstein2001unimodular,ng2001small,ellis2011trace,Josset:2016vrq} where the vacuum energy density of the quantum fields does not gravitate and the cosmological constant appears as an integration constant. There are several alternative formulations of unimodular gravity. Here the approach of Ref.~\cite{Bonder:2018mfz} is followed where the theory is described in the first-order formalism, with the internal group being LLT, and the so-called unimodular constraint is implemented through a Lagrange multiplier. Concretely, the gravitational action of unimodular Einstein--Cartan theory in $4$-dimensions is
\begin{align}
 \label{uecaction}
 S_{\rm UEC}\left[\omega^{ab},e^a,\mu; \varepsilon_0\right] &= \int\left[\frac{1}{2}\varepsilon_{abcd}R^{ab}\wedge e^c\wedge e^d + \mu\left(\varepsilon - \varepsilon_0\right) \right].
\end{align}
Here, lower case latin characters denote Lorentz indices, $\omega^{ab}$ and $e^a$ are the Lorentz connection and vielbein $1$-forms, $R^{ab}=\diff{}\omega^{ab} + \omega^{a}{}_c\wedge\omega^{cb}$ is the Lorentz curvature $2$-form, and Lorentz indices are raised and lowered with $\eta_{ab}$. In addition, $\mu=\mu(x)$ is a Lagrange multiplier $0$-form, $\varepsilon_{abcd}$ is the LLT-invariant totally-antisymmetric tensor such that $\varepsilon_{0123}=1$, $\varepsilon=\varepsilon_{abcd}e^a\wedge e^b \wedge e^c \wedge e^d/4!$ is the volume element $4$-form, and $\varepsilon_0$ is a nondynamical $4$-form.

It is easy to verify that $\varepsilon_0$ reduces the symmetries of the theory to Diff generated by vector fields $\xi$ restricted by the condition
\begin{align}\label{generalizeddiv}
\diff{i_\xi \varepsilon_0}=0.
\end{align}
On shell $\varepsilon_0=\varepsilon$, which implies that $\xi$ must be divergence free. Hence, this restricted set of Diff are usually known as volume preserving Diff (VPD). For concreteness, any Diff generated by a vector field satisfying Eq.~\eqref{generalizeddiv} is called a VPD.

An arbitrary variation of the action~\eqref{uecaction} gives
\begin{align}
 \delta S_{\rm UEC} &= \int\bigg( \delta\omega^{ab}\wedge\mathcal{E}_{ab}+\delta e^a\wedge \mathcal{E}_a + \delta\mu\,\mathcal{E} \bigg), 
 \end{align}
where 
\begin{align}
\mathcal{E}_{ab} &= \varepsilon_{abcd}T^c\wedge e^d,\\
 \mathcal{E}_a &= \varepsilon_{abcd}\left(R^{bc} - \frac{\mu}{3!}e^b\wedge e^c \right)\wedge e^d,\\
 \mathcal{E} &= \varepsilon - \varepsilon_0.
\end{align}
Here $T^a=\diff{}e^a+\omega^{a}{}_{b}\wedge e^b$ is the torsion $2$-form and all boundary terms are omitted. Now one must take the covariant derivative of these equations and try to write them as linear combinations of $ \mathcal{E}_{ab}$, $ \mathcal{E}_{a}$, and $ \mathcal{E}$; the result is
\begin{align}
\label{bianchiuec2}
\Diff{\mathcal{E}_{ab}} &= e_{[a}\wedge\mathcal{E}_{b]},\\
\label{bianchiuec1}
\Diff{\mathcal{E}_a} &=  i_a R^{bc}\wedge\mathcal{E}_{bc} +i_a T^b\wedge\mathcal{E}_b - i_a\diff{\mu}\,\mathcal{E} - \diff{\mu}\wedge i_a\varepsilon_0,\\
\label{bianchiuec3}
\Diff{\mathcal{E}} &= 0.
\end{align}
The next step is to contract Eqs.~\eqref{bianchiuec2} and \eqref{bianchiuec1} with the gauge parameters $\lambda^{ab}$ and $\xi^a$, respectively, and manipulate the left-hand side to get a total derivative. Integrating over spacetime, these equations can be brought into the form
\begin{align}
\label{bianchiuec2a}
0 &=\int\left[ \Diff{\lambda^{ab}} \wedge\mathcal{E}_{ab} - \lambda^{a}{}_b e^b \wedge\mathcal{E}_a\right],\\
\label{bianchiuec1a}
 0 &=\int\left[ i_\xi R^{ab} \wedge\mathcal{E}_{ab}+ \left(\Diff{\xi^a} + i_\xi T^a \right)\wedge\mathcal{E}_a  - i_\xi\diff{\mu}\,\mathcal{E}+ \mu\diff{i_\xi}\varepsilon_0\right] .
\end{align}
It is possible to verify that Eq.~\eqref{bianchiuec2a} has no symmetry breaking terms but the last term in Eq.~\eqref{bianchiuec1a} is a symmetry breaking term. Thus, the theory is invariant under LLT and the corresponding field transformations are
\begin{align} \label{LLT}
 \mbox{LLT} &= \begin{cases}
  \delta_{\rm LLT}(\epsilon) \omega^{ab} &= \Diff{\lambda^{ab}},\\
 \delta_{\rm LLT}(\epsilon) e^a &= -\lambda^{a}{}_b e^b,\\
 \delta_{\rm LLT}(\epsilon) \mu &= 0.
 \end{cases}
 \end{align}
 
In addition, one can conclude that the theory is not invariant under all LT, as expected from the fact that the theory only invariant under VPD. However, similarly to what occurs for the Diff, the theory is invariant under the subset of LT generated by vector fields satisfying Eq.~\eqref{generalizeddiv}, which are called volume preserving LT (VPLT). These VPLT act of the dynamical fields as
\begin{align} \label{VPLT}
 \mbox{VPLT} &= \begin{cases}
  \delta_{\rm VPLT}(\xi) \omega^{ab} &= i_\xi R^{ab},\\
 \delta_{\rm VPLT}(\xi) e^a &= \Diff{\xi^a} + i_\xi T^a,\\
 \delta_{\rm VPLT}(\xi) \mu &= i_\xi\diff{\mu},
 \end{cases}
 \end{align}
where $\xi$ satisfies Eq.~\eqref{generalizeddiv}.

At this point, the symmetries of the theory have been found. Therefore, one can turn to compute the algebra. It is important to stress that the results of Sec.~\ref{sec:LT} cannot be used because the gauge parameter is not arbitrary, as required by the method used to compute the algebra. Thus, one first needs to find an unrestricted parameter that generates the VPLT. Since all top forms are related by a proportionality factor, it is possible to write $\varepsilon_0 = e^\phi\varepsilon$, where $\phi$ is a scalar density transforming as $\delta_{\rm Diff}(\xi)\phi = \star\diff{\star}\xi$. In terms of this density, it is possible to locally solve Eq.~\eqref{generalizeddiv}, the solution is
\begin{align}\label{xisol}
 \xi = e^{-\phi} \star\diff{\alpha},
\end{align}
where $\star:\Omega^p\left(\mathcal{M}\right)\to\Omega^{N-p}\left(\mathcal{M}\right)$ is the Hodge dual and $\alpha$ is a $2$-form that plays the role of the unrestricted gauge parameter, which does not transform. Clearly, $\xi$ depends on the dynamical fields $\phi$ and $e^a$; the latter implicitly through the Hodge dual. Thus, under LLT, the gauge parameter subject to the condition~\eqref{generalizeddiv} transforms as
\begin{align}
 \delta_{\rm LLT}\left(\lambda\right)\xi^a = \mathring{\xi}^a - \lambda^{a}{}_b\xi^b,
\end{align}
where
\begin{align}\label{mathringxi}
 \mathring{\xi}^a = \frac{1}{2} e^{-\phi}\epsilon^{abcd}\left(D_b\tilde{\alpha}_{cd} + \tilde{\alpha}_{pd}T^{p}{}_{bc}\right),
\end{align}
with $\tilde{\alpha}_{ab}= \lambda^{c}{}_a\alpha_{cb} + \lambda^{c}{}_b\alpha_{ac}=-\tilde{\alpha}_{ba}$ and $T^{a}=\tfrac{1}{2} T^{a}{}_{bc}e^b\wedge e^c$.

Similarly, under VPLT, the vector field $\xi$ subject to the condition~\eqref{generalizeddiv} transforms as
\begin{align}
 \delta_{\rm VPLT}\left(\xi_1\right)\xi_2^a = \Lie_{\xi_1}\xi_2^a - \mathring{\xi}_2^a + \tilde{\lambda}_1^{a}{}_b\xi_2^b,
\end{align}
where $\tilde{\lambda}_1^{a}{}_b = i_{\xi_1}\omega^{a}{}_b$. Using these transformations, the commutators of LLT and VPLT, applied to either $e^a$, $\omega^{ab}$, and $\mu$, form the algebra
 \begin{align}
 \left[\delta_{\rm LLT}(\epsilon_1),\delta_{\rm LLT}(\epsilon_2)\right] &= \delta_{\rm LLT}(\lambda_3),\\
 \left[\delta_{\rm LLT}(\epsilon),\delta_{\rm VPLT}(\xi_1)\right] &= \delta_{\rm VPLT}(\mathring{\xi}), \\
 \left[\delta_{\rm VPLT}(\xi_1),\delta_{\rm VPLT}(\xi_2)\right] &= \delta_{\rm LLT}(\hat{\lambda}) + \delta_{\rm VPLT}(\hat{\xi}), 
 \end{align}
where $\lambda_{3}^{ab} =2\lambda_{1}^{[a}{}_c\lambda_{2}^{|c|b]}$, $\mathring{\xi}$ has been defined in Eq.~\eqref{mathringxi}, $\hat{\xi}^a = \Lie_{\xi_1}\xi_2^a + \mathring{\xi}_1^a - \mathring{\xi}_2^a$, and $\hat{\lambda}^{ab} = i_{\xi_1}i_{\xi_2}R^{ab}$, with $\xi_1$ and $\xi_2$ satisfying Eq.~\eqref{generalizeddiv}. Importantly, the algebra closes off shell and it only involves translations that are VPLT. This can be demonstrated by showing that $\mathring{\xi}$ and $\hat{\xi}$ satisfy the condition~\eqref{generalizeddiv} through $i_{\mathring{\xi}}\varepsilon_0 = -\diff{\tilde{\alpha}}$, where $\tilde{\alpha} = \tilde{\alpha}_{ab}e^b\wedge e^c/2$, and $i_{\hat{\xi}}\varepsilon_0 = \diff{}\left(i_{\xi_1}i_{\xi_2}\varepsilon_0 - \tilde{\alpha}_1 + \tilde{\alpha}_2 \right)$, where $\tilde{\alpha_i} = \left[\left(i_{\xi_i}\omega^{c}{}_a\right)\alpha_{cb} + \left(i_{\xi_i}\omega^{c}{}_b \right)\alpha_{ac}\right]e^b\wedge e^c/2$ with $i=1,2$. Clearly, it has the form of the algebra obtained in Sec.~\ref{sec:LT} but replacing LT by VPLT, which suggests that the algebra of Sec.~\ref{sec:LT} keeps its form even when the gauge parameters are restricted.

Finally, from the relation between LLT, Diff, and LT, it was expected that unimodular gravity would not be fully invariant under LT and/or LLT. Interestingly, the restriction on the Diff simply translates into a restriction on the LT; the invariance under LLT is unaffected. This is unexpected in view of the result that spontaneous LLT violation imply spontaneous violation of Diff and viceversa \cite{Nogo1,Nogo2,Nogo3}.

\section{Conclusions\label{sec:discussion}}

In the framework of gravity gauge theories, local translations can be thought of as improved diffeomorphisms. In this work it is shown that, for generic gauge theories of gravity, i.e., where the internal symmetry group, the spacetime dimensionality, the gravitational degrees of freedom, and the matter fields, are all arbitrary, the algebra composed by the local translations and the internal group closes off shell. Therefore, this result can be understood as a general feature of the gauge theories and not as a consequence of a particular assumption (e.g., local Lorentz invariance).

What is more, a method to find the symmetries in theories with explicit breaking of diffeomorphisms and/or internal symmetries is presented. Using the unimodular Einstein--Cartan theory as an example, it is verified that this method leads to the symmetries of the theory, and it shows that the form of the local translations is affected by the symmetry breaking. One important lesson of this work is that, to calculate the algebra, one first needs to solve the restriction on the gauge parameters and write them in terms of unrestricted parameters. When this is done, the algebra can be calculated in a straightforward way and it turns out that it still closes off shell. Moreover, even though the field transformations under local translations are sensitive to the symmetry breaking, the algebra does not change its structure. This should be regarded as a very strong evidence suggesting that the algebra of the local translations and the internal symmetry plays a fundamental role in the structure of any gauge theory of gravity.

The results of this work strengthen the proposal that gravity can be treated as a gauge theory of the local translations and the internal symmetry, which may have profound implications regarding the quantization of gravity. Still, very interesting questions remain open. For example, is the structure of a presumed gauge group that includes local translations obstructed by the Coleman--Mandula theorem~\cite{Coleman:1967ad}? Also, what should be the fundamental symmetries in the quantum realm? This could be particularly relevant in programs like loop quantum gravity where one of the major goals is to implement the Hamiltonian constraint.

\ack
The authors thank to O.~Castillo-Felisola, S.~del Pino, D.~Gonz\'alez, F.~Izaurieta, B.~Ju\'arez-Aubry, T.~Koslowski, M.~Montesinos, R.~Olea, J.~Oliva, and G.~Rubilar for their feedback. This work is supported by UNAM-DGAPA-PAPIIT Grant No.~RA101818 and UNAM-DGAPA postdoctoral fellowship.

\section*{References}

\bibliographystyle{unsrt}
\bibliography{References}

\end{document}